\documentclass[twocolumn,prc,a4paper,floatfix,nofootinbib,superscriptaddress]{revtex4-1}
\usepackage{amsmath,color,graphicx}
\usepackage{mathtools}
\usepackage{appendix}
\usepackage{subfigure}
\usepackage{color}
\usepackage{ulem}
\DeclareMathOperator{\Real}{Re}
\newcommand{\bra}[1]{\langle #1|}
\newcommand{\ket}[1]{|#1 \rangle}
\renewcommand{\Re}{\Real}
\newcommand{\HF}{\text{HF}}
\newcommand{\HFB}{\text{HFB}}

\newcommand{\eff}{\text{e}}
\newcommand{\FG}{\text{FG}}
\newcommand{\calE}{\mathcal{E}}
\newcommand{\Vantisymm}{\bar{V}}
\newcommand{\Vbraket}[4]{\bra{#1\;#2}\Vantisymm\ket{#3\;#4}}
\newcommand{\Vangleavg}{\Vantisymm_{\text{avg}}}
\renewcommand{\Vec}[1]{\mathbf{#1}}
\newcommand{\kf}{k_\text{F}}

\newcommand{\kv}{\Vec{k}}
\newcommand{\qv}{\Vec{q}}
\newcommand{\fm}{\text{fm}}
\newcommand{\fmi}{\text{fm}^{-1}}
\newcommand{\vlowk}{V_{\text{low-}k}}

\renewcommand{\Ref}[1]{Ref.~\cite{#1}}
\newcommand{\Refs}[1]{Refs.~\cite{#1}}
\newcommand{\Eq}[1]{Eq.~(\ref{#1})}
\newcommand{\Eqs}[1]{Eqs.~(\ref{#1})}
\newcommand{\Fig}[1]{Fig.~\ref{#1}}

\begin{document}
\title{Equation of state of superfluid neutron matter with low-momentum interactions}

\author{Viswanathan Palaniappan}
\email{viswanathan@physics.iitm.ac.in}
\affiliation{Department of Physics, Indian Institute of Technology
  Madras, Chennai - 600036, India}
\affiliation{Universit\'e Paris-Saclay, CNRS-IN2P3, IJCLab, 91405
  Orsay, France}
\author{S. Ramanan}
\email{suna@physics.iitm.ac.in}
\affiliation{Department of Physics, Indian Institute of Technology
  Madras, Chennai - 600036, India}
\author{Michael Urban}
\email{michael.urban@ijclab.in2p3.fr}
\affiliation{Universit\'e Paris-Saclay, CNRS-IN2P3, IJCLab, 91405
	Orsay, France}
%%%%%%%%%%%%%%%%%%%%%%%%%%%%%%%%%%%%%%%%%%%%%%%%%%%%%%%%%%%%%%%%%%%%%%%%%%%%%%%%%%
\begin{abstract} 
In this work, we calculate the ground state energy of pure neutron matter using
the renormalization group based low-momentum effective interaction $\vlowk$ in
Bogoliubov many-body perturbation theory (BMBPT), which is a perturbative
expansion around the Hartree-Fock-Bogoliubov (HFB) ground state. In order to
capture the low-density behavior of neutron matter, it turns out to be better to
use a density dependent cutoff in the $\vlowk$ interaction. Perturbative
corrections to the HFB energy up to third order are included. We find that at
low densities corresponding to the inner crust of neutron stars, the HFB state
that includes pairing is a better starting point for perturbation expansion. It
is observed that including the higher order perturbative corrections, the cutoff
dependence of the ground state energy is reduced.
\end{abstract}
\maketitle
%%%%%%%%%%%%%%%%%%%%%%%%%%%%%%%%%%%%%%%%%%%%%%%%%%%%%%%%%%%%%%%%%%%%%%%%%%%%%%%%%%
\section{Introduction}
%%%%%%%%%%%%%%%%%%%%%%%%%%%%%%%%%%%%%%%%%%%%%%%%%%%%%%%%%%%%%%%%%%%%%%%%%%%%%%%%%%
Neutron stars consist of several layers called the outer crust, inner crust,
outer core, and inner core. The outer crust is made of nuclei that get
progressively neutron rich, and a
degenerate electron gas to ensure charge neutrality~\cite{Chamel2008}. At some
point, the neutrons can no longer be bound, forming a gas of neutrons
interspersing a lattice of quasi-nuclei (clusters). This defines the inner crust.
The outer core consists of very neutron rich uniform nuclear matter, while the
composition of the inner core remains an open question. While the crust itself
extends just over $2$ km, understanding its structure is crucial to explain
certain observations. The
unbound neutrons in the inner crust are believed to be in a superfluid phase,
which is necessary to explain the observed pulsar glitches \cite{Anderson1975}. 
(However, for a detailed understanding of the pulsar glitches, one has to take into account 
the effect of the entrainment of the superfluid neutrons by the lattice of 
clusters \cite{Chamel2012, Andersson2012, Martin2016}
as well as the pinning of superfluid vortices \cite{Wlazlowski2016}.)
Furthermore, neutron pairing plays an important role in the thermal
evolution of the star \cite{Yakovlev2004}. While superfluidity in neutron-star 
crusts was predicted as early as 1960~\cite{Migdal1959}, a quantitatively 
reliable theoretical description still
eludes the community, largely due to the uncertainties in the input two-body
interaction and the subsequent medium and higher-body corrections. 

Pure neutron matter is a simple yet useful model for a neutron star. In first
approximation, it describes the gas of unbound neutrons in the inner crust,
neglecting the presence of the clusters. Furthermore, the properties of the inner
crust including the clusters depend sensitively on the equation of state of the
neutron gas, because the inner crust can be regarded as a phase coexistence of a
liquid and a gas phase \cite{Martin2015}. At low densities, corresponding to the
inner crust, the neutron-neutron ($nn$) interaction is attractive in the 
$^1\!S_0$ partial wave, resulting in the formation of spin-singlet Cooper pairs. 
At higher
densities, i.e. in the outer layers of the core, pairing occurs between neutrons
in the spin-triplet ($^3\!P_2-{}^3\!F_2$) channel. At the BCS level (i.e., 
free-space $nn$ interaction and single-particle spectrum), $^1\!S_0$ 
pairing is completely constrained by two-body scattering in free space. However, 
corrections beyond BCS are important, and the gap equation is very sensitive to 
such corrections. For example, it was seen in~\cite{Ramanan2020} that already the 
inclusion of the effective mass from different effective interactions in the 
single-particle 
energies introduces model dependence in the $^1\!S_0$ gaps. On the other hand, at 
the high densities as they are found in the outer core, the triplet channel gaps 
are highly model
dependent~\cite{Dong2013,Maurizio2014,Srinivas2016,Drischler2017,Papakonstantinou2017} and it 
is not surprising that a proper description of triplet pairing requires the 
inclusion of the three-nucleon force at the very least (see for 
example~\cite{Maurizio2014,Srinivas2016,Drischler2017,Papakonstantinou2017}).

Because of the fact that the $nn$ scattering length $a \approx -18\,\fm$ is much 
larger than the effective range $r_\eff \approx 2.7\,\fm$ of the $nn$ interaction,
it was suggested by Bertsch that dilute neutron matter could be in a 
first approximation modeled as a unitary Fermi gas (defined by $a\to\infty$ and
 $r_\eff\to 0$) \cite{Baker1999}. In the meanwhile, the experimental realization of 
the unitary Fermi gas and 
of the BCS-BEC crossover with ultracold trapped atoms has boosted also a lot of 
theoretical activity in this field, see \cite{CalvaneseStrinati2018} for a 
review. The present work 
extends our previous study for ultracold Fermi gases~\cite{Urban2021} to pure 
neutron matter. While ultracold atoms have tunable scattering length $a$ (via 
Feshbach resonance) and negligible effective range $r_\eff$, 
both these quantities are fixed in pure 
neutron matter by the $nn$ interaction. Only as long as the effects of the 
effective range can be neglected, both these systems exhibit universal 
behavior. In~\cite{Urban2021}, we studied the equation of state of a gas of 
ultracold fermions from the BCS regime ($1/\kf a\to -\infty$, where $\kf$ is the 
Fermi momentum) to the unitary limit $1/\kf a \to 0$), starting with the 
Hartree-Fock-Bogoliubov (HFB) energy, 
computed with a $\vlowk$ like interaction tailored for the cold atomic systems, 
and then including 
corrections to the HFB energy up to third order within the Bogoliubov many-body 
perturbation theory (BMBPT). In addition, we used the dependence of the results
on the cutoff of the effective low-momentum interaction to gain additional 
insights on the convergence of this scheme. 

The main focus of our current work is to adapt a strategy that was successfully 
used in nuclear structure calculations of finite nuclei \cite{Tichai2018}, 
to the case of infinite neutron matter. The aim is to obtain reliable results 
for the neutron-matter equation of state starting from a realistic $nn$ 
interaction, by 
performing essentially three steps: (1) The initial interaction is 
softened using renormalization-group (RG) techniques, lowering the momentum 
cutoff $\Lambda$ while keeping low-energy two-body observables 
unchanged \cite{Bogner2002,Bogner2003,Bogner2007,Bogner2010}.
(2) The resulting $\vlowk$ interaction is used in HFB approximation. (3) Corrections 
beyond HFB are included using BMBPT. In contrast to studies of finite nuclei,
since we are considering uniform 
matter, we can use a density dependent cutoff $\Lambda = f \kf$, where $f$ is a 
scale factor. Such a scaling was used for the first time in \cite{Schwenk2003}.
Previous studies \cite{Urban2021,Ramanan2018} have demonstrated that a variable cutoff is 
especially important to describe the low-density limit.
Since $\Lambda$ is an unphysical parameter, any dependence of the
equation of state on $\Lambda$ or on the scale factor $f$ gives an indication for 
the importance of missing three-body and medium corrections.

Our study is similar in spirit to the work by Coraggio et 
al.~\cite{Coraggio2013}, who compared results obtained with interactions having 
different regulator functions, using the Hartree-Fock (HF) instead of the HFB 
approximation as a starting point and third-order many-body perturbation theory 
(MBPT). They showed that the dependence on the choice of the 
regulator is to a large extent compensated if in addition to the two-body also 
three-body interactions are included. However, that work focused on higher 
densities where pairing effects on the equation of state are weak, and where the 
use of a density dependent cutoff is not required. If one is interested in 
superfluidity, it is of course mandatory to start from the HFB and not from the
HF ground state.

The paper is organized as follows. Sect.~\ref{sec:formalism} discusses the BMBPT 
formalism and obtains expressions for the second and third order corrections to 
the HFB energy. The main results are discussed in Sect.~\ref{sec:results}, while 
in Sect.~\ref{sec:conclusion}, we summarize the important aspects of our current 
work and look at possible directions that could be explored in future studies.
%%%%%%%%%%%%%%%%%%%%%%%%%%%%%%%%%%%%%%%%%%%%%%%%%%%%%%%%%%%%%%%%%%%%%%%%%%%%%%%%%%
\section{Formalism} \label{sec:formalism}
%%%%%%%%%%%%%%%%%%%%%%%%%%%%%%%%%%%%%%%%%%%%%%%%%%%%%%%%%%%%%%%%%%%%%%%%%%%%%%%%%%
In this section, we outline the BMBPT, by first reviewing the HFB approach to 
incorporate the superfluid nature of the ground state. 
%%%%%%%%%%%%%%%%%%%%%%%%%%%%%%%%%%%%%%%%%%%%%%%%%%%%%%%%%%%%%%%%%%%%%%%%%%%%%%%%%%
\subsection{Hartree-Fock-Bogoliubov Theory}\label{sec:HFB}
%%%%%%%%%%%%%%%%%%%%%%%%%%%%%%%%%%%%%%%%%%%%%%%%%%%%%%%%%%%%%%%%%%%%%%%%%%%%%%%%%%
The pairing between two particles in the states $\kv\uparrow$ and
$-\kv\downarrow$ in an interacting Fermi gas can be realized via the definition 
of a new quasiparticle operator~\cite{FetterWalecka}
\begin{equation}
	\beta^{}_{\kv\,\sigma} = u^{}_k \, a^{}_{\kv\,\sigma} -
	(-1)^{\frac{1}{2}-\sigma} \, v^{}_k \, a^{\dagger}_{-\kv\,-\sigma}\,,
\end{equation}
where the coefficients $u^{}_k$ and $v^{}_k$ can be chosen to be real, 
$a^{}_{\kv\,\sigma}$ and $a^{\dagger}_{\kv\,\sigma}$ are 
particle annihilation and creation operators, and $\sigma = \pm \tfrac{1}{2}$ 
labels the spin projection. For the transformation to be canonical, the 
quasiparticle operators have to satisfy the usual anticommutation relations and
as a result, the coefficients are constrained to obey the condition 
$u_k^2 + v_k^2 =1$. Since the particle number is not conserved in this approach, 
one fixes the average particle number (or number density $n$) by introducing the
chemical potential $\mu$ and writing the grand canonical Hamiltonian as
\begin{align}
	\hat{K} =& \hat{H} - \mu \hat{N}\nonumber\\
	=& \sum_{\kv\sigma} (\varepsilon_k^0 - \mu)\, 
	a^{\dagger}_{\kv\,\sigma} a^{}_{\kv\,\sigma}\nonumber\\
	 &+ \frac{1}{4} \sum_{\kv^{}_i \sigma^{}_i} 
	\Vbraket{\kv^{}_1\sigma^{}_1}{\kv^{}_2\sigma^{}_2}{\kv^{}_3\sigma^{}_3}{\kv^{}_4\sigma^{}_4}\nonumber\\ 
	&\qquad\qquad\times a^{\dagger}_{\kv^{}_1\sigma^{}_1} 
	a^{\dagger}_{\kv^{}_2\sigma^{}_2}
	a^{}_{\kv^{}_4\sigma^{}_4} a^{}_{\kv^{}_3\sigma^{}_3}\,.
	\label{eq:H_gc}
\end{align}  
The second sum is over all momenta (with the constraint of momentum conservation) 
and spins. Since we are considering infinite matter, the sums over momenta will 
actually be integrals as in \Ref{Urban2021}. Further, $\varepsilon_k^0 = k^2/2m$
denotes the energy of a free neutron with mass $m$, $\hat{N}$ is the particle-number
operator, and $\Vantisymm$ is the antisymmetrized potential whose matrix elements are
given in terms of those in the partial wave basis by
\begin{multline}
\Vbraket{\kv^{}_1 \sigma^{}_1}{\kv^{}_2 \sigma^{}_2}{\kv^{}_3 \sigma^{}_3}{\kv^{}_4 \sigma^{}_4}\\
 = (4\pi)^2 \sum_{sm^{}_sm'_s} \sum_{ll' m^{}_l m'_l} \sum_{jm^{}_j}
 C^{sm^{}_s}_{\frac{1}{2} \sigma^{}_1\frac{1}{2}\sigma^{}_2}
 C^{sm'_s}_{\frac{1}{2} \sigma^{}_3\frac{1}{2}\sigma^{}_4}\\
  \times C^{jm^{}_j}_{l m^{}_l sm^{}_s} C^{jm^{}_j}_{l' m'_l sm'_s} \, 
  Y^{}_{lm^{}_l}(\hat{{\qv}}_{12})  Y^*_{l'm'_l}(\hat{{\qv}}_{34})\\ 
  \times i^{l-l'} \, V^{}_{jll's} (q_{12},{q}_{34}) \; \big[1+(-1)^{l+s}\big].
  \label{Vpartialwaves}
\end{multline}
where $\qv_{ij} = (\kv_i - \kv_j)/2$, $C^{j m}_{j_1 m_1 j_2 m_2}$
are the Clebsch-Gordan coefficients in the notation of \Ref{Varshalovich} and 
$Y_{l m}$ the spherical harmonics.
Rewriting the expression for the Hamiltonian in terms of the quasiparticle 
operators, \Eq{eq:H_gc} becomes
\begin{multline}\label{eq:HFB_hamiltonian}
\hat{K} =
  \sum_{\kv} \big[ 2 v_k^2 \xi^{}_k - v_k^2 \; \Sigma_{\HFB}(k) - u^{}_k 
	v^{}_k \; \Delta^{}_k \big] \\
 + \sum_{\kv,\sigma} \beta^{\dagger}_{\kv\,\sigma} \, \beta^{}_{\kv\,\sigma} \; 
 \big[ (u_k^2 - v_k^2) \,\xi^{}_k + 2 u^{}_k v^{}_k \Delta^{}_k \big] \\
	+ \frac{1}{2} \sum_{\kv,\sigma} (-1)^{\frac{1}{2}-\sigma} \big[
	\beta^{}_{-\kv\,-\sigma} \, \beta^{}_{\kv\,\sigma} + 
	\beta^{\dagger}_{\kv\,\sigma} \, \beta^{\dagger}_{-\kv\,-\sigma} \big]\\
   \times \big[ 2 u^{}_k v^{}_k \,\xi^{}_k - (u_k^2 - v_k^2) \Delta^{}_k 
    \big] + \mathcal{N}(\hat{V}),
\end{multline}
where $\Sigma_{\HFB}({k})$ and $\Delta_k$ are, respectively, the HFB self-energy
and the $^1\!S_0$ gap function whose expressions will be given below and 
$\xi^{}_{k}$ denotes the single-particle energy measured with respect to
the chemical potential,
\begin{align}
	\xi^{}_k \equiv \varepsilon_k^0 + \Sigma_{\HFB}(k) - \mu.
	\label{eq:xi_hfb}
\end{align}
The symbol $\mathcal{N}$ in \Eq{eq:HFB_hamiltonian} denotes normal ordering with respect to the quasiparticle 
operators (moving $\beta^\dagger$ operators to the left and $\beta$ operators to 
the right), and $\mathcal{N}(\hat{V})$ is given by
\begin{multline}
	\mathcal{N}(\hat{V}) =
	  \frac{1}{4} \sum_{\kv^{}_i\sigma^{}_i}
	\Vbraket{\kv^{}_1\sigma^{}_1}{\kv^{}_2\sigma^{}_2}{\kv^{}_3\sigma^{}_3}{\kv^{}_4\sigma^{}_4}\\
   \times \mathcal{N}\big(a^{\dagger}_{\kv^{}_1\sigma^{}_1} 
 a^{\dagger}_{\kv^{}_2\sigma^{}_2} a^{}_{\kv^{}_4\sigma^{}_4} 
 a^{}_{\kv^{}_3\sigma^{}_3}\big).
\end{multline}
As usual, one requires the third term in \Eq{eq:HFB_hamiltonian} to 
vanish, which leads to
\begin{equation}\label{eq:uv}
	u^{}_k v^{}_k = \frac{\Delta^{}_k}{2E^{}_k}\,,\quad
 	u_k^2 = \frac{1}{2} \bigg(1+\frac{\xi^{}_k}{E^{}_k} \bigg),
 \quad
	v_k^2 = \frac{1}{2} \bigg(1-\frac{\xi^{}_k}{E^{}_k} \bigg),
\end{equation}
where
\begin{equation}
    E^{}_k \equiv \sqrt{\Delta_k^2+\xi_k^2}
    \label{eq:E_qp}
\end{equation}
is the quasiparticle energy. The $^1\!S_0$ gap and HFB self-energy are given by
\begin{gather}
	\Delta^{}_{k} = - \frac{1}{\pi} \int dk' \, k'^2 \; 
	\frac{\Delta^{}_{k'}}{E^{}_{k'}} \; V^{}_{^1\!S_0}(k,k'),
	\label{eq:gap_1s0}\\
\Sigma_{\HFB}(k)
= \frac{1}{\pi} \int dk' \, k'^{\,2} \, v_{k'}^2 \Vangleavg(k,k').
\label{eq:HFB_selfenergy}
\end{gather}
In \Eq{eq:HFB_selfenergy}, $\Vangleavg(k,k')$ denotes the angle averaged 
interaction
\begin{multline}
\Vangleavg(k,k') = \frac{1}{2} \int d(\cos\theta_{\kv',\kv}) 
  \sum_{slj} (2j+1)
\\
  \times  V^{}_{sllj}(q,q) \; \big[1+(-1)^{l+s}\big],
\end{multline}
where $\qv = (\kv-\kv')/2$. In the limit of zero gap,
the factor $v_k^2$ becomes the Heaviside $\theta$ function and 
\Eq{eq:HFB_selfenergy} 
reduces to the familiar HF self-energy. 
For a given chemical potential, the number density in the HFB approximation is 
given by
\begin{equation}
	n^{}_{\HFB} = \frac{1}{\pi^2} \int dk \,k^2\, v_k^2 \, ,
	\label{eq:nhfb}
\end{equation}
and the HFB ground state energy density is given by
\begin{equation}
	\calE_{\HFB} = \frac{1}{2\pi^2} \int dk \,k^2\, 
	\bigg[v_k^2\,(k^2+\Sigma_{\HFB}(k)) - \frac{\Delta_k^2}{E^{}_k}\bigg].
	\label{eq:hfb_gs}
\end{equation}
The energy per particle $E^{}_\HFB$ is $\calE^{}_{\HFB}/n^{}_{\HFB}$. 
%%%%%%%%%%%%%%%%%%%%%%%%%%%%%%%%%%%%%%%%%%%%%%%%%%%%%%%%%%%%%%%%%%%%%%%%%%%%%%%%%%
\subsection{Bogoliubov Many-Body Perturbation Theory}\label{sec:BMBPT}
%%%%%%%%%%%%%%%%%%%%%%%%%%%%%%%%%%%%%%%%%%%%%%%%%%%%%%%%%%%%%%%%%%%%%%%%%%%%%%%%%%
With \Eq{eq:uv}, the operator $\hat{K}$ of \Eq{eq:HFB_hamiltonian} can now be 
rewritten as
\begin{equation}
	\hat{K} = K_{00} + \hat{K}_{11} + \mathcal{N}(\hat{V}),
\end{equation}
where $\hat{K}_{ij}$ contains $i$ quasiparticle creation operators and $j$ 
quasiparticle annihilation operators. For example, $K_{00}$ and $\hat{K}_{11}$ 
are given by
\begin{align}
	K_{00} &= \sum_{\kv} \bigg( v_k^2 \big[2\xi^{}_{k}-\Sigma_{\HFB}(k)\big] - 
	\frac{\Delta_k^2}{2E^{}_k} \bigg),
	\\
	\hat{K}_{11} &= \sum_{\kv,\sigma} \beta^{\dagger}_{\kv\,\sigma} \, 
	\beta^{}_{\kv\,\sigma} \; E^{}_k\,,
\end{align}
where $K_{00}$ corresponds to the expectation value of the operator $\hat{K}$ in 
the HFB ground state which has zero quasiparticles, and $\hat{K}_{11}$ describes 
the energy of non-interacting quasiparticles. The interaction between 
quasiparticles is contained in $\mathcal{N}(\hat{V})$, which can be written as 
\begin{equation}
	\mathcal{N}(\hat{V})= \hat{K}_{40} + \hat{K}_{31} + \hat{K}_{22} + 
	\hat{K}_{13} + \hat{K}_{04}\,.
\end{equation}
Since the eigenstates and eigenvalues of $\hat{K}_0 = K_{00} + \hat{K}_{11}$ are 
known, one can build corrections to the HFB ground state through perturbation 
theory, i.e., by writing
\begin{equation}
\hat{K} = \hat{K}_0 + \lambda\, \mathcal{N}(\hat{V})
\end{equation}
and expanding the ground state of $\hat{K}$ in powers of the formal parameter 
$\lambda$, the physical situation corresponding of course to $\lambda=1$.
Following \cite{Tichai2018,Urban2021}, there is no first-order correction and the
second-order (BMBPT2) correction to the ground-state energy density%
\footnote{\label{note_grandpotential} As mentioned in \Ref{Urban2021},
for a given chemical potential $\mu$, the second- and third-order corrections 
to the grand potential $\Omega = \calE-\mu n$ coincide
with the second- and third-order corrections to the energy density of the 
system with the density $n_{\HFB}(\mu)$.}
is given by
\begin{align}
\calE^{(2)} = - \frac{1}{4!} \sum_{1234} 
  \frac{|\bra{0}\hat{K}_{04}\ket{1234} |^2}{E_{1234}} \,,
\label{eq:sec_order}
\end{align}
where $|0\rangle$ is the HFB ground state, the indices 1, 2, 3, and 4 mean both 
momentum and spin, e.g., $1 = \{\kv_1,\sigma_1\}$, such that the summation over 
$1234$ means summation over momenta $\kv_1\dots\kv_4$ (with zero total momentum 
$\kv_1+\kv_2+\kv_3+\kv_4=0$) and spins $\sigma_1\dots\sigma_4$. The energy of
the intermediate four-quasiparticle state $\ket{1234} = 
\beta^{\dagger}_1\beta^{\dagger}_2\beta^{\dagger}_3\beta^{\dagger}_{4} \ket{0}$ 
is given by $E_{1234} = E_{k_1} + E_{k_2} + E_{k_3} + E_{k_4}$. The factor $4!$
accounts for the number of permutations of indices 1234 describing all the same 
state. The explicit form of the operator $\hat{K}_{04}$ in \Eq{eq:sec_order} is 
\begin{multline}
\hat{K}_{04} = 
 -\frac{1}{4} \sum_{1234} \Vbraket{1}{2}{3}{4}\; 
(-1)^{\sigma_1 + \sigma_2} \\
  \times v_1 v_2 u_4 u_3\,\beta_{-1}\,\beta_{-2}\,\beta_4\,\beta_3\,.
\end{multline}
where $v_{1}=v_{k_{1}}$, etc. With a bit of algebra, it can be shown that the
second-order correction to the energy is given by
\begin{equation}
\calE^{(2)} = - \sum_{\kv^{}_1\kv^{}_2\kv^{}_3} \frac{A + B + C}
  {E_{1234}},
\end{equation}
where 
\begin{align}\label{BMBPT2A}
A =& \frac{1}{4} \, \,v_{1}^2 v_{2}^2 u_{3}^2 u_{4}^2  
  \sum_{\sigma_1\sigma_2\sigma_3\sigma_4} |\Vbraket{-1}{-2}{3}{4}|^2, \\
\label{BMBPT2B}
B =& \frac{1}{4} \, u_{1}v_{2}\,u_{2}v_{2}\,u_{3}v_{3}\,u_{4}v_{4}
  \sum_{\sigma_1\sigma_2\sigma_3\sigma_4} |\Vbraket{-1}{-2}{3}{4}|^2, \\
\label{BMBPT2C}
C =& v_{1}^2\,u_{2}v_{2}\,u_{3}v_{3}\,u_{4}^2 
  \sum_{\sigma_1\sigma_2\sigma_3\sigma_4} (-1)^{\sigma_2 + \sigma_3} 
  \nonumber \\
&\times \Re\Big[\Vbraket{-1}{-2}{3}{4}^* \Vbraket{-1}{-3}{2}{4} \Big].
\end{align}
The terms $B$ and $C$ contribute only when there is a finite gap. In the 
limit of no pairing, only term $A$ contributes, and one retrieves the
HF+MBPT2 as already noted in~\cite{Urban2021}.
In deriving \Eqs{BMBPT2A}-(\ref{BMBPT2C}), the hermiticity
\begin{equation}
\Vbraket{1}{2}{3}{4} = \Vbraket{3}{4}{1}{2}^*,
\end{equation}
and the time-reversal symmetry
\begin{equation}
\Vbraket{1}{2}{3}{4} = (-1)^{\sigma_1+\sigma_2+\sigma_3+\sigma_4} \Vbraket{-3}{-4}{-1}{-2},
\end{equation}
have been used.

Similarly, the third-order (BMBPT3) correction to the ground state
energy density is given by
\begin{equation}{\label{eq:bmbpt3}}
\calE^{(3)} = \sum_{1\dots8} \frac{ 
  \bra{0}\hat{K}_{04}\ket{1234} 
  \bra{1234}\hat{K}_{22}\ket{5678}
  \bra{5678}\hat{K}_{40}\ket{0}}
{E_{1234}E_{5678}},
\end{equation}
where $\hat{K}^{}_{40} = \hat{K}_{04}^\dagger$ and
\begin{align}\label{eq:operator_K22}
	\hat{K}_{22} 
	=& \frac{1}{4} \sum_{1234} \bra{1 \; 2} \Vantisymm \ket{3 \; 4} \; \big[ 
	u_1 u_2 u_4 u_3\,\beta^{\dagger}_1\beta^{\dagger}_2\beta_4\beta_3
    \nonumber \\
	&+ (-1)^{\sigma_1 + \sigma_2 + \sigma_3 + \sigma_4} v_1 v_2 v_4 v_3 \,
	\beta^{\dagger}_{-4}\beta^{\dagger}_{-3}\beta_{-1}\beta_{-2} \nonumber \\ 
	& - (-1)^{\sigma_2 + \sigma_3} u_1 v_2 u_4 v_3 \,
	\beta^{\dagger}_1\beta^{\dagger}_{-3}\beta_{-2}\beta_4 \nonumber \\
	&+ (-1)^{\sigma_2 + \sigma_4} u_1 v_2 v_4 u_3\,
	\beta^{\dagger}_1\beta^{\dagger}_{-4}\beta_{-2}\beta_3 \nonumber \\
	& + (-1)^{\sigma_1 + \sigma_3} v_1 u_2 u_4 v_3 \,
	\beta^{\dagger}_2\beta^{\dagger}_{-3}\beta_{-1}\beta_4 \nonumber \\
	&- (-1)^{\sigma_1 + \sigma_4} v_1 u_2 v_4 u_3 \,
	\beta^{\dagger}_2\beta^{\dagger}_{-4}\beta_{-1}\beta_3 \big].
\end{align}
For ease of presentation, the lengthy explicit expressions at third order
(that were obtained with the help of a Mathematica code)
are relegated to the appendix. Momentum conservation finally
reduces the number of momentum integrations to four. 

The momentum integrals are computed numerically using Monte-Carlo integration
with the importance-sampling method explained in \Ref{Urban2021}.
%%%%%%%%%%%%%%%%%%%%%%%%%%%%%%%%%%%%%%%%%%%%%%%%%%%%%%%%%%%%%%%%%%%%%%%%%%%%%%%%%%
\section{Results and discussion} \label{sec:results}
%%%%%%%%%%%%%%%%%%%%%%%%%%%%%%%%%%%%%%%%%%%%%%%%%%%%%%%%%%%%%%%%%%%%%%%%%%%%%%%%%%
\subsection{HFB results}
%%%%%%%%%%%%%%%%%%%%%%%%%%%%%%%%%%%%%%%%%%%%%%%%%%%%%%%%%%%%%%%%%%%%%%%%%%%%%%%%%%
Before we present our results, let us say a few words about the interaction used
in our computations. We use the free-space RG-based low-momentum interaction 
$\vlowk$ (with a smooth exponential regulator with $n_{\text{exp}} = 5$), which
depends on the renormalization cutoff $\Lambda$ \cite{Bogner2007}. Lowering the 
cutoff through the RG flow softens the 
interaction while preserving two-body observables by construction. Furthermore, 
for cutoffs below $\Lambda\lesssim 2.1\,\fmi$, the $\vlowk$ matrix elements 
become practically independent of the choice of the initial interaction such as 
AV18 or chiral potentials \cite{Bogner2010}. Here we will show results obtained 
with the $\vlowk$ derived from AV18 \cite{Wiringa1995}, but when starting from 
the N3LO chiral interaction \cite{Entem2003} we obtain very similar results.

When such soft interactions are used in a many-body calculation, one hopes that 
perturbative corrections show rapid convergence. The two-body
interaction eventually flows to the two-body scattering length as 
$\Lambda \to 0$. Therefore, small cutoffs become very important for describing 
physics at low densities (which otherwise would require ladder resummations), as 
was observed in~\cite{Ramanan2018}. There, a density dependent cutoff 
$\Lambda = f \,\kf$, with a scale factor $f$ of the order of 2.5 (sufficiently
large to leave the BCS gap unchanged), was required to reproduce the 
Gor'kov-Melik-Bharkhudarov results for the superfluid transition 
temperature when screening effects are included. In a subsequent study 
\cite{Urban2021}, a $\vlowk$ like interaction was used to describe Fermi gases
with contact interactions, and at least for $\kf|a| \ll 1$, convergence of the
HFB+BMBPT3 scheme was reached for cutoffs in the range
$\Lambda \lesssim 2.5\,\kf$.

Given \Eq{Vpartialwaves}, which is an
expansion in partial waves, it is important
to investigate the number of such waves that need to be included in the calculation
of the equation of state.
%%%%%%%%%%%%%%%%%%%%%%%%%%%%%%%%%%%%%%%%%%%%%%%%%%%%%%%%%%%%%%%%%%%%%%%%%%%%%%%%%%
\begin{figure}[h]
  %convergence partial waves
  % \includegraphics[width=\columnwidth]{convergence_fixed_cutoff.eps}\\
  \includegraphics[width=\columnwidth]{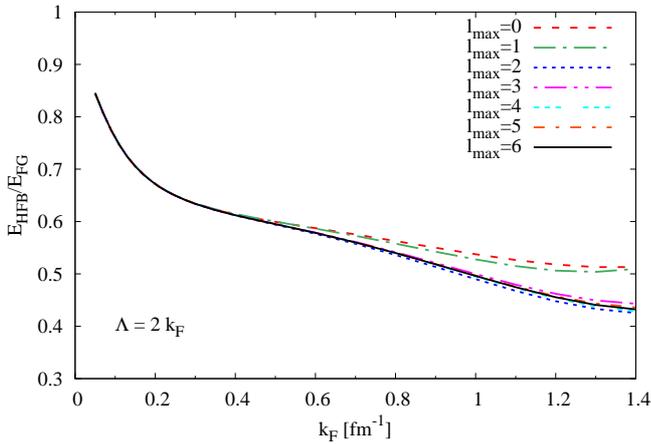}
  \caption{Convergence of the equation of state (energy density in units of the
    energy density of the ideal Fermi gas as a function of $\kf$) in the HFB 
    approximation with the inclusion of the higher partial waves. Here,
    $l_{\max}$ indicates the highest partial wave that has been included.}
    \label{fig:convergence_cutoff}
\end{figure}
%%%%%%%%%%%%%%%%%%%%%%%%%%%%%%%%%%%%%%%%%%%%%%%%%%%%%%%%%%%%%%%%%%%%%%%%%%%%%%%%%%
\Fig{fig:convergence_cutoff} shows the HFB ground-state energy density [see 
\Eq{eq:hfb_gs}] in units of the energy density of the non-interacting Fermi gas
(FG),
\begin{equation}
\calE_{\FG} = \frac{\kf^5}{10\pi^2m}
\end{equation}
as a function of $\kf = (3\pi^2 n)^{1/3}$, as various partial waves are included 
one by one, for a density dependent cutoff $\Lambda = 2\,\kf$. We see that the 
$S$-wave dominates until
$\kf \sim 0.4 \, \fmi$, beyond which the $P$ and $D$ waves become important. 
Because of strong cancellations between the $^3\!P_0$, $^3\!P_1$ and $^3\!P_2$ 
contributions, including the $P$ wave without the $D$ wave 
does not result in a noticeable improvement at any density. 
Partial waves beyond $l = 2$ have some effect for $\kf \gtrsim 0.7\, 
\fmi$. We conclude that for the density range we are interested in, including 
partial waves with $l\leq 6$ yields converged (with respect to $l_{\max}$) 
results.

Let us now come back to the discussion of constant vs.\ density dependent cutoffs.
%%%%%%%%%%%%%%%%%%%%%%%%%%%%%%%%%%%%%%%%%%%%%%%%%%%%%%%%%%%%%%%%%%%%%%%%%%%%%%%%%%
\begin{figure}[h]
    \includegraphics[width=\columnwidth]{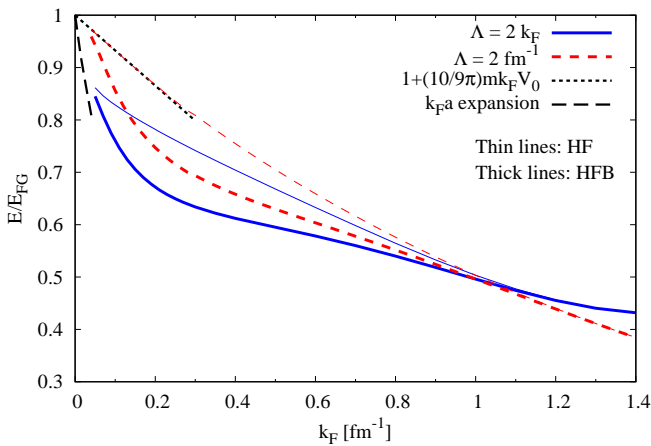}
    \caption{HF (thin lines) and HFB (thick lines) results for the ground-state 
    energy
    obtained with a fixed cutoff $\Lambda = 2\,\fmi$ (red dashes) and with a
    density-dependent cutoff $\Lambda = 2\kf$ (blue solid lines). The short black
    dashes show the asymptotic low-density 
    behavior of the HF result with fixed cutoff according to \Eq{eq:HF_lowkf}, 
    while the long black dashes show the 
    fourth-order $\kf a$ expansion of \Ref{Wellenhofer2021}.}
    \label{fig:hf_vs_hfb}
\end{figure}
%%%%%%%%%%%%%%%%%%%%%%%%%%%%%%%%%%%%%%%%%%%%%%%%%%%%%%%%%%%%%%%%%%%%%%%%%%%%%%%%%%
In \Fig{fig:hf_vs_hfb}, we compare HF (thin lines) and HFB (thick lines) results
obtained with fixed ($\Lambda = 2\,\fmi$, red dashes) and density dependent
($\Lambda = 2\kf$, blue solid lines) cutoffs. It is evident that results for the 
constant and the density dependent cutoffs agree at $\kf = 1 \, 
\fmi$. For larger $\kf$, the density dependent cutoff is bigger than the fixed 
cutoff, reaching eventually a value of $2.8 \, \fmi$ (since the figure covers the 
range $\kf\le 1.4\,\fmi$). We observe that at $\kf\gtrsim 1.1\,\fmi$, the HFB 
results are practically identical to the respective HF results because the pairing
gap becomes very small compared to the Fermi energy.

As the density tends towards zero, the ground state
energy of the interacting system approaches that of the non-interacting system.
In HF approximation, for a fixed cutoff, one can obtain an expansion in $\kf$,
\begin{equation}
    \frac{E_\HF}{E_\FG} = 1 + \frac{10}{9\pi} m \kf V_0\,, 
    \label{eq:HF_lowkf}
\end{equation}
where $V_0 = V_{^1\!S_0}(0,0)$ is the matrix element of the two-body interaction
for $q = q' = 0$. This result, shown as black short-dashed line in 
\Fig{fig:hf_vs_hfb} is in perfect agreement with the HF result (thin red dashes) 
up to $\kf\sim 0.2\,\fmi$. The inclusion of pairing in HFB does not change this
asymptotic behavior since the pairing gap vanishes as $\kf^2 e^{-\pi/|2\kf a|}$ 
\cite{CalvaneseStrinati2018}, and this explains why also the HFB
results (thick red dashes) eventually approach the curve given by
\Eq{eq:HF_lowkf} at very small $\kf$.

Obviously, \Eq{eq:HF_lowkf} is in disagreement with the well-known leading term
of the $\kf a$ expansion \cite{FetterWalecka} which is given by \Eq{eq:HF_lowkf}
with the replacement $V_0\to a/m$. In the case of neutron 
matter and $\Lambda = 2\,\fmi$, the factor $a/m$ is about ten times larger in
magnitude than $V_0$. Hence, the slope of the HF (and HFB) results at small
$\kf$ with fixed $\Lambda = 2\,\fmi$ is much too small, as can be seen by 
comparing them with the results of the $\kf a$ expansion of 
\Ref{Wellenhofer2021} [black dashes, including orders up to $(\kf a)^4$].
Notice that the validity of this expansion 
is limited to the small region $\kf \lesssim 1/{|a|} \sim 0.05 \,\fmi$. Furthermore, 
it does not include effects of the finite range of the $nn$ interaction.

It is well known that with decreasing cutoff, $V_0$ grows in
magnitude until it finally approaches $a/m$ (see, e.g., Fig.~15 
in \Ref{Bogner2010}). The HF(B) energies obtained with the density dependent 
cutoff (solid blue lines in \Fig{fig:hf_vs_hfb}) are therefore much lower
(at low densities) than those obtained with the fixed cutoff, and they
are in much better agreement with the
$\kf a$ expansion. We conclude that, in order to reproduce the correct
low-density behavior, the HFB approximation with a density dependent cutoff
is a better starting point than with a fixed cutoff.

%%%%%%%%%%%%%%%%%%%%%%%%%%%%%%%%%%%%%%%%%%%%%%%%%%%%%%%%%%%%%%%%%%%%%%%%%%%%%%%%%%
   \begin{figure}[h]
    \centering
    \includegraphics[width=\columnwidth]{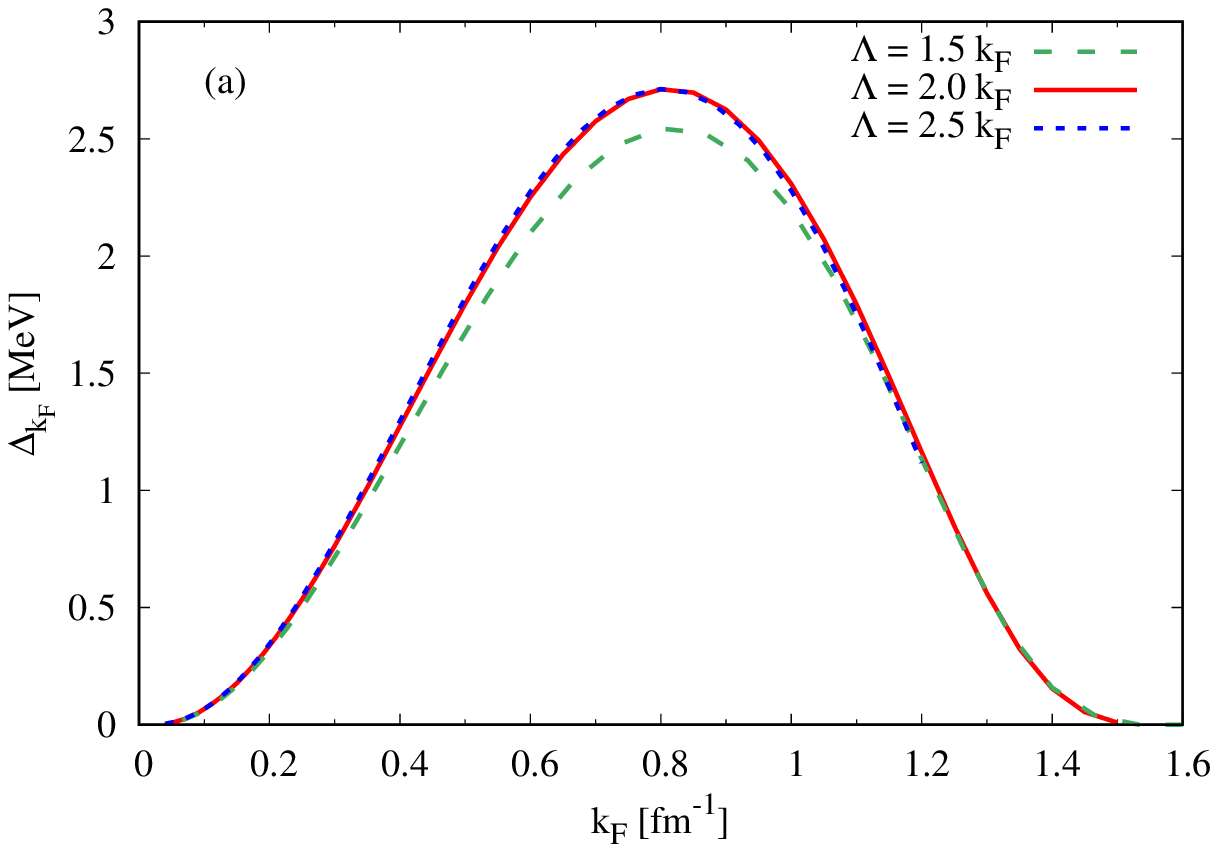}
    \includegraphics[width=\columnwidth]{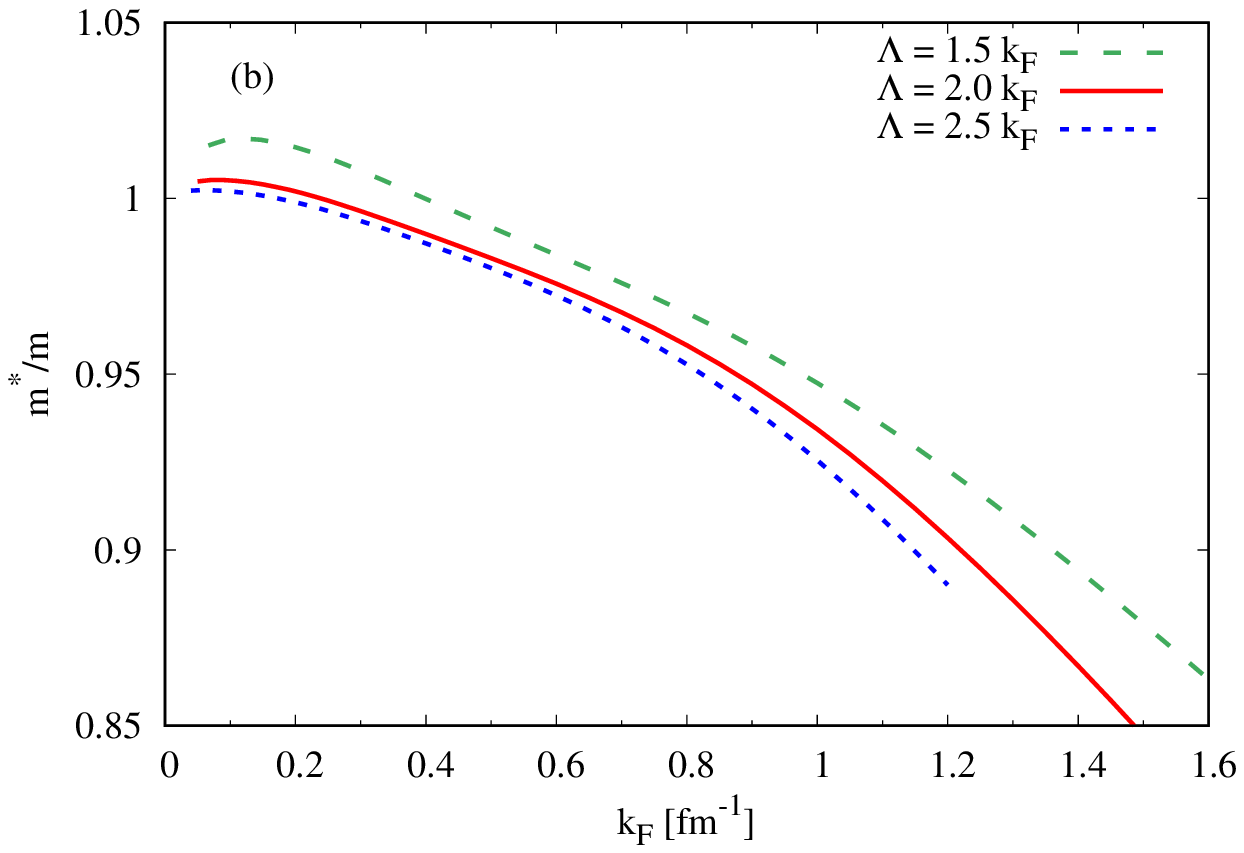}
    \caption{(a) HFB $^1\!S_0$ pairing gap and (b) 
    effective mass as a function of $\kf$ for three different values of the
    cutoff parameter $f = \Lambda/\kf$.
    \label{fig:hfb_results}}
\end{figure} 
%%%%%%%%%%%%%%%%%%%%%%%%%%%%%%%%%%%%%%%%%%%%%%%%%%%%%%%%%%%%%%%%%%%%%%%%%%%%%%%%%%
The HFB $^1\!S_0$ gap $\Delta_{\kf}$ as function of $\kf$, computed using
\Eqs{eq:xi_hfb} and (\ref{eq:E_qp})-(\ref{eq:nhfb}) for different values of the
cutoff parameter $f=\Lambda/\kf$, is seen in \Fig{fig:hfb_results}(a), while
\Fig{fig:hfb_results}(b) shows the 
corresponding effective mass, defined by
\begin{equation}
\frac{1}{m^*} = 
  \frac{1}{m}+\frac{1}{\kf}\frac{d\Sigma_{\HFB}(k)}{dk}\Big|_{k=\kf}.    
\end{equation}
The unusual fact that $m^*>m$ at very low density, especially for $\Lambda = 
1.5\,\kf$, can be understood from the shape of the matrix elements of $\vlowk$
in the case of very small cutoffs, see, e.g., Fig.~11 of \Ref{Ramanan2018}. 
Since the effective mass determines the density of states at the Fermi level, the
gap usually reacts very sensitively to its change, a reduction of $m^*$ leading 
to a reduction of the gap. But in \Fig{fig:hfb_results} we note that, while the 
effective mass decreases with increasing cutoff parameter, the dependence of the 
gap on this parameter shows the opposite trend. The explanation is that if the
cutoff gets very small ($\Lambda = 1.5\, \kf$), the matrix elements of the 
potential and hence the gap function $\Delta_k$ and the $v_k$ factors are cut off 
closely above $\kf$ so that the gap must be reduced, even though the effective
mass is enhanced. For $\Lambda/\kf$ between 2 and 2.5, these two effects 
compensate each other (the almost exact cancellation being accidental), and 
the gap remains practically unchanged in spite of the further reduced
effective mass.

%%%%%%%%%%%%%%%%%%%%%%%%%%%%%%%%%%%%%%%%%%%%%%%%%%%%%%%%%%%%%%%%%%%%%%%%%%%%%%%%%%
\subsection{HFB+BMBPT results}
%%%%%%%%%%%%%%%%%%%%%%%%%%%%%%%%%%%%%%%%%%%%%%%%%%%%%%%%%%%%%%%%%%%%%%%%%%%%%%%%%%
So far, we have only discussed HFB results and their dependence on the unphysical 
cutoff parameter. Let us now discuss how the situation improves when we include
the BMBPT corrections.
%%%%%%%%%%%%%%%%%%%%%%%%%%%%%%%%%%%%%%%%%%%%%%%%%%%%%%%%%%%%%%%%%%%%%%%%%%%%%%%%%%
\begin{figure}[h]
    \includegraphics[width=\columnwidth]{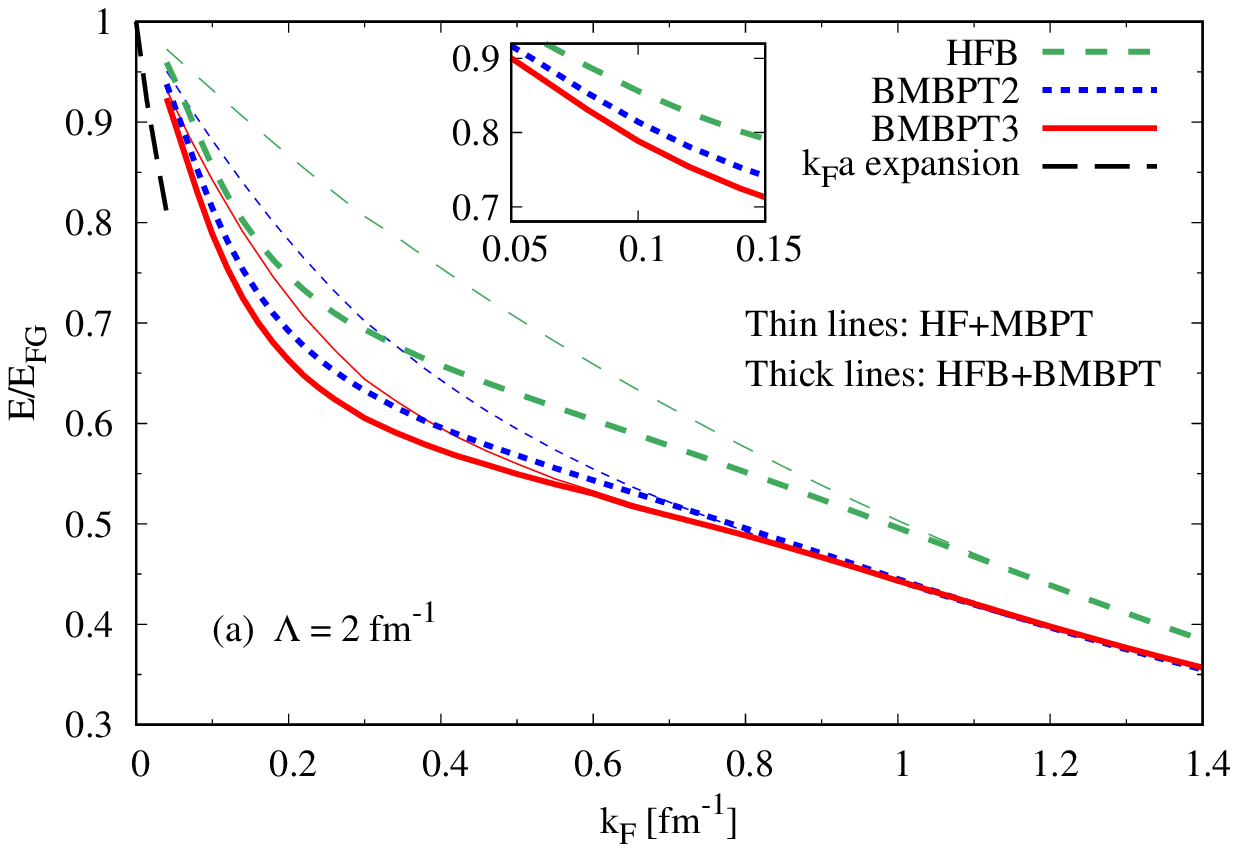}
    \includegraphics[width=\columnwidth]{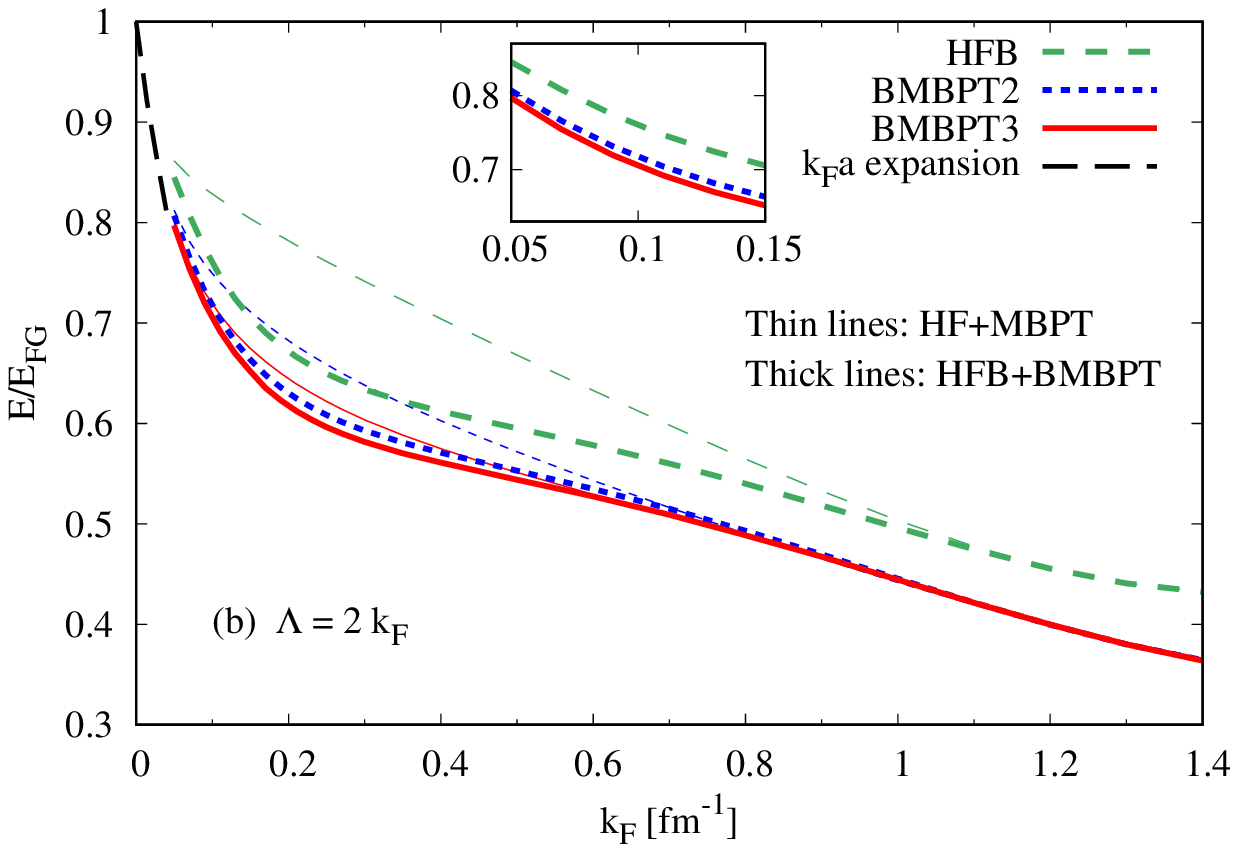}
    \caption{Ground state energy in units of the energy of the free Fermi gas, as
    a function of $\kf$, at different levels of approximation: HFB (green dashes), BMBPT2 (short blue dashes), BMBPT3 (red solid lines), for (a) fixed cutoff
    $\Lambda = 2\,\fmi$ and (b) density dependent cutoff $\Lambda = 2\,\kf$.
    HF and MBPT results are shown using thin lines.
    \label{fig:pt_results}}

\end{figure}
%%%%%%%%%%%%%%%%%%%%%%%%%%%%%%%%%%%%%%%%%%%%%%%%%%%%%%%%%%%%%%%%%%%%%%%%%%%%%%%%%%
In Fig.~\ref{fig:pt_results}, we present our  
calculation of the equation of state within HFB+BMBPT up to 
third order for both fixed cutoff [$\Lambda = 2\,\fmi$, 
Fig.~\ref{fig:pt_results}(a)] and density dependent cutoff [$\Lambda = 2\kf$, 
Fig.~\ref{fig:pt_results}(b)] as a function of $\kf$. In addition, we also 
include the HF+MBPT results obtained by setting 
$\Delta=0$. 

The BMBPT3 results obtained with fixed
and density dependent cutoffs differ noticeably from each
other in the region $\kf\lesssim 0.4\,\fmi$. In particular, the energies obtained with a fixed cutoff (\Fig{fig:pt_results}(a))
are much higher than those of the $\kf a$ expansion. Furthermore, the BMBPT3 correction, e.g., at $\kf=0.1\,\fmi$, is about 60\,\% of 
the BMBPT2 correction (see inset in \Fig{fig:pt_results}(a)), so one cannot claim that the expansion has converged, and 
it is not clear whether perturbation theory will ever be able to bring the
results down to the $\kf a$ expansion. On the contrary, with the density dependent cutoff (\Fig{fig:pt_results}(b)) the
agreement between BMBPT3 and $\kf a$ expansion is very good, and the BMBPT3 brings only a tiny correction to the BMBPT2 result, so
that one may speak of convergence of the BMBPT expansion. 

Let us explain these findings. At very low density, the results should agree with the 
$\kf a $-expansion because the particles scatter at very low energies and 
Pauli-blocking effects are negligible.
As already pointed out in the preceding subsection, in order to obtain the scattering length, one either
has to resum ladders to all orders or take the limit $\Lambda \to 0$ in which case the value of 
$V_0=V_{^1S_0}(0,0)$ approaches the scattering length. Hence the HFB energy with a fixed cutoff is 
far too high, as already seen in \Fig{fig:hf_vs_hfb},
and the third-order perturbation theory is not enough to 
correct for that,\footnote{This can also be understood by looking at
the so-called Weinberg eigenvalues as shown in Fig.~1 of \Ref{Ramanan2007}: although $\Lambda=2\,\fmi$ is enough
to make the interaction soft, in the sense that all repulsive eigenvalues are small, 
the attractive eigenvalue becomes large at low density and hence very high orders in perturbation theory
would be needed to describe the full (in-medium) $T$ matrix.} 
whereas for density dependent cutoff, at very low
$\kf$ (and hence very low $\Lambda$), the perturbative corrections bring the results into agreement
with the $\kf a$-expansion.

At very low and high densities, the gaps become very small and HFB+BMBPT reduces 
in practice to the simpler HF+MBPT, plotted with thin lines. For 
$\kf\lesssim 1 \, \fmi$, where the gap is large enough to
make a noticeable contribution to the equation of state, we see that the perturbative 
corrections within BMBPT to HFB are much lesser than those of 
MBPT to HF. Hence the HFB+BMBPT 
converges better than the HF+MBPT as expected. However, it is 
interesting that in the region $\kf\sim 0.6-1\,\fmi$, both MBPT3 and BMBPT3 give practically
the same results, i.e., the MBPT is able to account for the pairing 
correlations which are missing in the HF ground state. For 
$\kf\lesssim 0.6\,\fmi$, the MBPT cannot reproduce the BMBPT results, at least 
not yet at 3rd order.

%%%%%%%%%%%%%%%%%%%%%%%%%%%%%%%%%%%%%%%%%%%%%%%%%%%%%%%%%%%%%%%%%%%%%%%%%%%%%%%%%%
\begin{figure}[h]
    \includegraphics[width=\columnwidth]{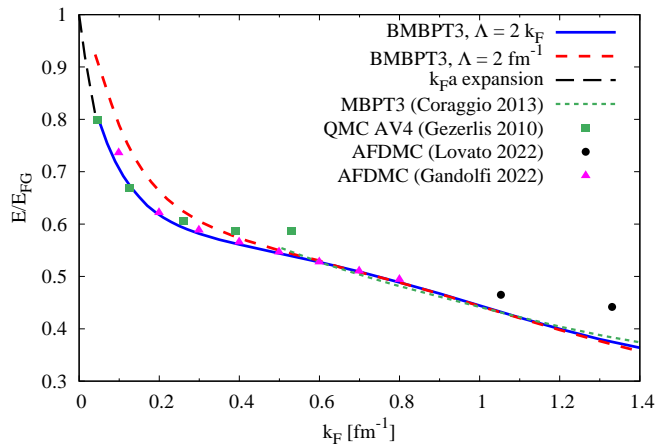}
    \caption{BMBPT3 equation of state for fixed cutoff $\Lambda = 2\,\fmi$ (red
    dashes) and density dependent cutoff $\Lambda = 2\,\kf$ (blue 
    solid lines). For comparison, we also show the low-density behavior according
    to the fourth-order $\kf a$ expansion of \Ref{Wellenhofer2021} 
    (black long dashes), MBPT3 results of \Ref{Coraggio2013} 
    (green small dashes), and various QMC results of \Refs{Gezerlis2010}
    (green squares), \cite{Gandolfi2022} (purple triangles), and \cite{Lovato2022}
    (black circles).
    \label{fig:bmbpt3_results}}
\end{figure}
%%%%%%%%%%%%%%%%%%%%%%%%%%%%%%%%%%%%%%%%%%%%%%%%%%%%%%%%%%%%%%%%%%%%%%%%%%%%%%%%%%
In Fig.~\ref{fig:bmbpt3_results}, we compare our final results (BMBPT3) with
results from the literature such as MBPT \cite{Coraggio2013} and Quantum Monte
Carlo (QMC) \cite{Gezerlis2010, Lovato2022, Gandolfi2022}. At low densities, our results for $\Lambda = 2\kf$ are in
excellent agreement with the QMC results
\cite{Gezerlis2010,Gandolfi2022} [except the last two points of
\Ref{Gezerlis2010} (green squares) at $\kf\sim 0.4-0.5\,\fmi$]. Futhermore,
our results are also very similar to the MBPT results of \Ref{Coraggio2013} (green 
small dashes), although another interaction (chiral
N3LO with $\Lambda=500$ MeV, no three-body force) was used 
there.

Looking at the region $\kf>1\,\fmi$, we have seen in
\Fig{fig:hf_vs_hfb} that the HFB results (which are close to the HF results in
that region) obtained with $\Lambda=2\,\fmi$ differ from those obtained with 
$\Lambda = 2\kf$. But now we see in \Fig{fig:bmbpt3_results} that this difference
is to a large extent absorbed by the BMBPT corrections, as it should be since
physical results should be cutoff independent. But we also observe that our
results lie clearly below the QMC points of \Ref{Lovato2022} (black circles). 
This difference could be due to the missing three-body force in our calculation.

%%%%%%%%%%%%%%%%%%%%%%%%%%%%%%%%%%%%%%%%%%%%%%%%%%%%%%%%%%%%%%%%%%%%%%%%%%%%%%%%%%
%Cutoff dependence
\begin{figure}[h]
    \centering
    \includegraphics[width=\columnwidth]{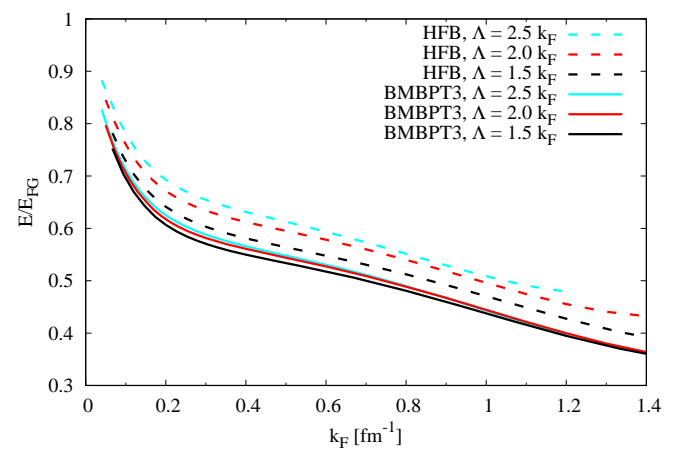}
    \caption{HFB (dashed lines) and BMBPT3 (solid lines)
    equations of state for three different scale factors 
    $f=\Lambda/\kf = 1.5$ (black), 2 (red), and 2.5 (light blue).}
    \label{fig:cutoff_dependence}
\end{figure}
%%%%%%%%%%%%%%%%%%%%%%%%%%%%%%%%%%%%%%%%%%%%%%%%%%%%%%%%%%%%%%%%%%%%%%%%%%%%%%%%%%
In Fig.~\ref{fig:cutoff_dependence}, we study more systematically the cutoff 
dependence of the HFB (dashed 
lines) and BMBPT3 (solid lines) results by using different scale parameters $f = 
\Lambda/\kf$ (as in \Fig{fig:hfb_results}). Again, we note that, compared to the 
HFB results, the cutoff dependence is significantly reduced by including
perturbative 
corrections up to third order. But the equation state including the third-order
corrections still does have residual cutoff dependence. This could be due to 
missing higher-order perturbative contributions, or missing contributions of
three-body forces, as discussed in the following section. 

%%%%%%%%%%%%%%%%%%%%%%%%%%%%%%%%%%%%%%%%%%%%%%%%%%%%%%%%%%%%%%%%%%%%%%%%%%%%%%%%%%
\section{Summary and outlook} \label{sec:conclusion}
%%%%%%%%%%%%%%%%%%%%%%%%%%%%%%%%%%%%%%%%%%%%%%%%%%%%%%%%%%%%%%%%%%%%%%%%%%%%%%%%%%
In this work, we focus on calculating the equation of state for pure neutron 
matter with renormalization-group softened $\vlowk$ interactions within BMBPT
that builds perturbative corrections around the HFB state and hence takes
into account the superfluid nature of the ground state. While pairing almost 
does not affect the equation of state at high densities, including it certainly
gives a better starting  point for the perturbation theory in the range of
densities where the gap is large and where the HFB energy is considerably lower
than the HF one. Furthermore, at very low densities, the HFB provides a better
starting point of the perturbative expansion if one uses in the $\vlowk$ 
interaction a density dependent cutoff $\Lambda$ that scales with $\kf$ instead 
of a fixed cutoff.

Since the cutoff $\Lambda$ is an unphysical quantity, observables such as the
ground-state energy should not depend on it. While the BMBPT corrections
absorb the cutoff dependence of the HFB results to a large extent,
we note that there is still residual cutoff dependence after including third
order. For a cutoff of $\Lambda$ = 2 $\fmi$,
the three-body forces contribute at leading order only for 
$\kf > 0.8\, \fmi$~\cite{Hebeler2010}. However, for a density dependent cutoff
that scales with $\kf$, three- (and maybe even higher-) body forces could also
play a significant role at smaller densities because then the cutoff becomes
very small and the three- (and higher-) body terms may grow as the two-body 
term is evolved via the RG flow \cite{Nogga2004,Bogner2010}. At the very least,
it would be necessary to include the induced three-body
effects. An interesting approach to include the induced forces is the in-medium
SRG (IMSRG) technique that has been very successful in finite nuclei (see for example~\cite{Hergert2016}). 
However, this is beyond the scope of our current work and will be postponed
to a future study.

Furthermore, there is also a lot of interest in the superfluid properties of
neutron matter, independently of the equation of state. Like the HFB
ground-state energy, also the pairing gap shows dependence
on the unphysical cutoff parameter $\Lambda$ of the $\vlowk$ interaction. 
One may hope that, by computing higher-order perturbative corrections to the
normal and anomalous self-energies (i.e., diagonal and non-diagonal in 
Nambu-Gor'kov indices \cite{Schrieffer}), one can obtain more cutoff
independent results for the quasiparticle dispersion relation (mean field,
effective mass) and for the gap. Higher-order contributions to the anomalous
self-energy include, e.g., screening corrections to the gap, which are known
to reduce the gap. Work in this direction is in progress.

%%%%%%%%%%%%%%%%%%%%%%%%%%%%%%%%%%%%%%%%%%%%%%%%%%%%%%%%%%%%%%%%%%%%%%%%%%%%%%%%%%
\begin{acknowledgments} 
We thank L. Coraggio for sending us the data of \Ref{Coraggio2013}.
We acknowledge support from the Collaborative Research Program of IFCPAR/CEFIPRA,
Project number: 6304-4.
\end{acknowledgments}
\appendix
\section*{Appendix: Explicit expressions for the third-order BMBPT correction}
%\label{sec:appendix}
For completeness, we give here the expressions necessary to
compute $\Omega^{(3)}$, \Eq{eq:bmbpt3}, generalizing the
expressions given in the appendix of \Ref{Urban2021} to the
case of non-separable interactions with partial waves beyond
the $^1\!S_0$. 

Momentum conservation in $\hat{K}_{04}$ and $\hat{K}_{40}$ requires
$\kv_1+\kv_2+\kv_3+\kv_4=0$ and $\kv_5+\kv_6+\kv_7+\kv_8=0$. 
Furthermore, $\hat{K}_{22}$ leaves two momenta and two spins 
unchanged. So, there are only four independent momenta
to integrate and six spins to sum over. After relabeling the 
particles such that the four integration variables are called
$\kv_1\dots\kv_4$, we can
write
\begin{multline}
\label{eq:bmbpt3b}
\Omega_0^{(3)} 
= \sum_{\kv^{}_1\dots\kv^{}_4} \Big( 
u_{1}v_{1}\,u_{2}v_{2}\,u_{3}v_{3}\,u_{4}v_{4}\,A_1 \\
+ u_{1}v_{1}\,u_{2}v_{2}\,u_{3}v_{3}\,v^2_{4}\,A_2 
+ u_{1}v_{1}\,u_{2}v_{2}\,u_{3}v_{3}\,u^2_{4}\,A_3 \\
+ u_{1}v_{1}\,u_{2}v_{2}\,v^2_{3}\,v^2_{4}\,A_4
+ u_{1}v_{1}\,u_{2}v_{2}\,v^2_{3}\,u^2_{4}\,A_5 \\
+ v^2_{1}\,v^2_{2}\,v^2_{3}\,u^2_{4}\,A_6
+ v^2_{1}\,v^2_{2}\,u^2_{3}\,u^2_{4}\,A_7\Big).
\end{multline}
The remaining momenta are then given by different linear 
combinations of $\kv_1\dots\kv_4$ which we denote by
\begin{align}
  \kv_5 &= -\kv_1-\kv_2-\kv_3\,,&
  \kv_6 &= -\kv_1-\kv_2-\kv_4\,,\nonumber\\
  \kv_7 &= -\kv_1-\kv_3-\kv_4\,,&
  \kv_8 &= -\kv_2-\kv_3-\kv_4\,,\nonumber\\
  \kv_9 &= \kv_1+\kv_2-\kv_3\,,&
  \kv_{10} &= \kv_1+\kv_3-\kv_2\,,\nonumber\\
  \kv_{11} &= \kv_1+\kv_2-\kv_4\,,&
  \kv_{12} &= \kv_1+\kv_4-\kv_2\,,\nonumber\\
  \kv_{13} &= \kv_2+\kv_3-\kv_4\,.
\end{align}
Using notations
\begin{gather}
    \Vbraket{\kv_1\sigma_1}{\kv_2\sigma_2}{\kv_3\sigma_3}{\kv_4\sigma_4} = \Vantisymm_{\sigma_1\sigma_2\sigma_3\sigma_4}(\qv_{1,2},\qv_{3,4})
    \nonumber\\
    \qv_{i,j} = \frac{\kv_i-\kv_j}{2}\,\quad \qv^+_{i,j} = \frac{\kv_i+\kv_j}{2}\,,\quad \bar{\sigma}_i = -\sigma_i\,,\nonumber
\end{gather}
the expressions for the integrands $A_1\dots A_7$ in 
\Eq{eq:bmbpt3b} read:
\begin{widetext}
\begin{multline}
    A_1 = \sum_{\sigma_1\dots\sigma_6} \Bigg[ 
    (-1)^{\sigma_3+\sigma_6}\;u^2_{5}\,u^2_{6}\; \frac{\Re\big[\Vantisymm_{\sigma_2\sigma_4\bar{\sigma}_6\bar{\sigma}_1}(\qv_{2,4},\qv_{1,6}) \, \Vantisymm_{\bar{\sigma}_3\bar{\sigma}_1\sigma_2\sigma_5}(\qv_{1,3},\qv_{2,5}) \, \Vantisymm_{\sigma_3\sigma_5\sigma_4\sigma_6}(\qv_{3,5},\qv_{4,6})\big]}{E_{1,2,3,5}\;E_{1,2,4,6}}
    \\
    + (-1)^{\sigma_3+\sigma_6}\;v^2_{5}\,v^2_{6}\; \frac{\Re\big[\Vantisymm_{\sigma_1\sigma_4\bar{\sigma}_6\bar{\sigma}_2}(\qv_{1,4},\qv_{2,6}) \, \Vantisymm_{\bar{\sigma}_3\bar{\sigma}_2\sigma_1\sigma_5}(\qv_{2,3},\qv_{1,5}) \, \Vantisymm_{\sigma_3\sigma_5\sigma_4\sigma_6}(\qv_{3,5},\qv_{4,6})\big]}{E_{1,2,3,5}\;E_{1,2,4,6}}
    \\
    -(-1)^{\sigma_2+\sigma_3+\sigma_5+\sigma_6}\;(u^2_{5}\,u^2_{13}+v^2_{5}\,v^2_{13})\; \frac{\Re\big[\Vantisymm_{\sigma_1\sigma_4\bar{\sigma}_5\bar{\sigma}_6}(\qv_{1,4},\qv_{13,5}) \, \Vantisymm_{\sigma_2\sigma_3\sigma_6\sigma_4}(\qv_{2,3},\qv_{13,4}) \, \Vantisymm_{\bar{\sigma}_3\bar{\sigma}_2\sigma_1\sigma_5}(\qv_{2,3},\qv_{1,5})\big]}{E_{1,4,5,13}\;E_{1,2,3,5}}
    \\
    - (-1)^{\sigma_3+\sigma_5}\;u_{5}v_{5}\,u_{6}v_{6}\; \frac{\Re\big[\Vantisymm_{\sigma_2\sigma_4\bar{\sigma}_6\bar{\sigma}_1}(\qv_{2,4},\qv_{1,6}) \, \Vantisymm_{\bar{\sigma}_3\bar{\sigma}_1\sigma_2\sigma_5}(\qv_{1,3},\qv_{2,5}) \, \Vantisymm_{\sigma_3\bar{\sigma}_6\sigma_4\bar{\sigma}_5}(\qv^+_{3,6},\qv^+_{4,5})\big]}{E_{1,2,3,5}\;E_{1,2,4,6}}
    \\
    + 2\, (-1)^{\sigma_1+\sigma_6}\;(u^2_{8}\,v^2_{7}+u^2_{7}\,v^2_{8})\; \frac{\Re\big[\Vantisymm_{\sigma_1\sigma_4\bar{\sigma}_5\bar{\sigma}_3}(\qv_{1,4},\qv_{3,7}) \, \Vantisymm_{\bar{\sigma}_3\sigma_2\sigma_4\bar{\sigma}_6}(\qv_{2,3},\qv_{4,8}) \, \Vantisymm_{\bar{\sigma}_5\bar{\sigma}_1\sigma_6\sigma_2}(\qv_{1,7},\qv_{2,8})\big]}{E_{1,2,7,8}\;E_{1,3,4,7}}
    \\
    + (-1)^{\sigma_1+\sigma_2+\sigma_3+\sigma_4}\;u^2_{6}\,v^2_{5}\; \frac{\Re\big[\Vantisymm_{\sigma_1\sigma_2\bar{\sigma}_6\bar{\sigma}_4}(\qv_{1,2},\qv_{4,6}) \, \Vantisymm_{\bar{\sigma}_2\bar{\sigma}_1\sigma_3\sigma_5}(\qv_{1,2},\qv_{3,5}) \, \Vantisymm_{\sigma_5\bar{\sigma}_6\sigma_4\bar{\sigma}_3}(\qv^+_{5,6},\qv^+_{3,4})\big]}{E_{1,2,3,5}\;E_{1,2,4,6}}
    \Bigg],
\end{multline}

\begin{multline}
    A_2 = \sum_{\sigma_1\dots\sigma_6} \Bigg[ 
    u^2_{6}\,u_{9}v_{9}\; \frac{\Re\big[\Vantisymm_{\sigma_2\bar{\sigma}_5\sigma_3\bar{\sigma}_1}(\qv^+_{2,9},\qv^+_{1,3}) \, \Vantisymm_{\sigma_3\sigma_6\bar{\sigma}_5\bar{\sigma}_4}(\qv_{3,6},\qv_{4,9}) \, \Vantisymm_{\bar{\sigma}_4\bar{\sigma}_1\sigma_2\sigma_6}(\qv_{1,4},\qv_{2,6})\big]}{E_{1,2,4,6}\;E_{3,4,6,9}}
    \\
    - 2\,(-1)^{\sigma_1+\sigma_2+\sigma_4+\sigma_6}\;u^2_{6}\,u_{5}v_{5}\; \frac{\Re\big[\Vantisymm_{\sigma_2\sigma_3\bar{\sigma}_5\bar{\sigma}_1}(\qv_{2,3},\qv_{1,5}) \, \Vantisymm_{\bar{\sigma}_4\bar{\sigma}_2\sigma_1\sigma_6}(\qv_{2,4},\qv_{1,6}) \, \Vantisymm_{\sigma_4\bar{\sigma}_5\sigma_3\bar{\sigma}_6}(\qv^+_{4,5},\qv^+_{3,6})\big]}{E_{1,2,3,5}\;E_{1,2,4,6}}
    \\
    + \frac{1}{8} (-1)^{\sigma_1+\sigma_2+\sigma_3+\sigma_5}\;u_{9}v_{9}\,v^2_{6}\; \frac{\Re\big[\Vantisymm_{\sigma_1\sigma_2\sigma_3\sigma_5}(\qv_{1,2},\qv_{3,9}) \, \Vantisymm_{\bar{\sigma}_2\bar{\sigma}_1\sigma_4\sigma_6}(\qv_{1,2},\qv_{4,6}) \, \Vantisymm_{\sigma_4\sigma_6\bar{\sigma}_5\bar{\sigma}_3}(\qv_{4,6},\qv_{3,9})\big]}{E_{1,2,4,6}\;E_{3,4,6,9}}
    \\
     + \frac{1}{4} (-1)^{\sigma_1+\sigma_2+\sigma_4+\sigma_6}\;u_{5}v_{5}\,v^2_{6}\; \frac{\Re\big[\Vantisymm_{\sigma_1\sigma_2\bar{\sigma}_5\bar{\sigma}_3}(\qv_{1,2},\qv_{3,5}) \, \Vantisymm_{\bar{\sigma}_2\bar{\sigma}_1\sigma_4\sigma_6}(\qv_{1,2},\qv_{4,6}) \, \Vantisymm_{\bar{\sigma}_5\bar{\sigma}_3\bar{\sigma}_6\bar{\sigma}_4}(\qv_{3,5},\qv_{4,6})\big]}{E_{1,2,3,5}\;E_{1,2,4,6}}
     \Bigg],  
\end{multline}

\begin{multline}
    A_3 = \sum_{\sigma_1\dots\sigma_6} \Bigg[\frac{1}{4} (-1)^{\sigma_1+\sigma_2+\sigma_4+\sigma_6}\;u^2_{6}\,u_{5}v_{5}\; \frac{\Re\big[\Vantisymm_{\sigma_1\sigma_2\bar{\sigma}_6\bar{\sigma}_4}(\qv_{1,2},\qv_{4,6}) \, \Vantisymm_{\bar{\sigma}_2\bar{\sigma}_1\sigma_3\sigma_5}(\qv_{1,2},\qv_{3,5}) \, \Vantisymm_{\sigma_3\sigma_5\sigma_4\sigma_6}(\qv_{3,5},\qv_{4,6})\big]}{E_{1,2,3,5}\;E_{1,2,4,6}}
    \\
    + \frac{1}{8}(-1)^{\sigma_1+\sigma_2+\sigma_3+\sigma_5}\;u^2_{6}\,u_{9}v_{9}\; \frac{\Re\big[\Vantisymm_{\sigma_1\sigma_2\sigma_3\sigma_5}(\qv_{1,2},\qv_{3,9}) \, \Vantisymm_{\bar{\sigma}_2\bar{\sigma}_1\sigma_4\sigma_6}(\qv_{1,2},\qv_{4,6}) \, \Vantisymm_{\sigma_4\sigma_6\bar{\sigma}_5\bar{\sigma}_3}(\qv_{4,6},\qv_{3,9})\big]}{E_{1,2,4,6}\;E_{3,4,6,9}}
    \Bigg],
\end{multline}

\begin{multline}
    A_4 = \sum_{\sigma_1\dots\sigma_6} \Bigg[ -2 \, (-1)^{\sigma_3+\sigma_5}\;u^2_{13}\,u^2_{5}\; \frac{\Re\big[\Vantisymm_{\sigma_1\sigma_4\bar{\sigma}_5\bar{\sigma}_6}(\qv_{1,4},\qv_{13,5}) \, \Vantisymm_{\bar{\sigma}_3\bar{\sigma}_2\sigma_1\sigma_5}(\qv_{2,3},\qv_{1,5}) \, \Vantisymm_{\sigma_3\bar{\sigma}_6\sigma_4\bar{\sigma}_2}(\qv^+_{13,3},\qv^+_{2,4})\big]}{E_{1,4,5,13}\;E_{1,2,3,5}}
    \\
    - (-1)^{\sigma_3+\sigma_5}\;u^2_{5}\,u^2_{6}\; \frac{\Re\big[\Vantisymm_{\sigma_2\sigma_4\bar{\sigma}_6\bar{\sigma}_1}(\qv_{2,4},\qv_{1,6}) \, \Vantisymm_{\bar{\sigma}_3\bar{\sigma}_1\sigma_2\sigma_5}(\qv_{1,3},\qv_{2,5}) \, \Vantisymm_{\sigma_3\bar{\sigma}_6\sigma_4\bar{\sigma}_5}(\qv^+_{3,6},\qv^+_{4,5})\big]}{E_{1,2,3,5}\;E_{1,2,4,6}}
    \\
    - (-1)^{\sigma_2+\sigma_3+\sigma_4+\sigma_6}\;u^2_{7}\,v^2_{8}\; \frac{\Re\big[\Vantisymm_{\sigma_1\sigma_5\bar{\sigma}_2\bar{\sigma}_6}(\qv^+_{1,8},\qv^+_{2,7}) \, \Vantisymm_{\sigma_3\sigma_4\sigma_5\sigma_2}(\qv_{3,4},\qv_{2,8}) \, \Vantisymm_{\bar{\sigma}_4\bar{\sigma}_3\sigma_1\sigma_6}(\qv_{3,4},\qv_{1,7})\big]}{E_{1,2,7,8}\;E_{1,3,4,7}}
    \\
    + 2\, (-1)^{\sigma_1+\sigma_2+\sigma_3+\sigma_4+\sigma_5+\sigma_6}\;u^2_{11}\,u^2_{5}\; \frac{\Re\big[\Vantisymm_{\sigma_2\sigma_3\bar{\sigma}_5\bar{\sigma}_1}(\qv_{2,3},\qv_{1,5}) \, \Vantisymm_{\bar{\sigma}_4\bar{\sigma}_3\sigma_6\sigma_5}(\qv_{3,4},\qv_{11,5}) \, \Vantisymm_{\sigma_4\bar{\sigma}_2\sigma_1\bar{\sigma}_6}(\qv^+_{2,4},\qv^+_{1,11})\big]}{E_{1,2,3,5}\;E_{3,4,5,11}}
    \\
    - \frac{1}{2}(-1)^{\sigma_1+\sigma_3+\sigma_4+\sigma_6}\;u^2_{7}\,u^2_{8}\; \frac{\Re\big[\Vantisymm_{\sigma_3\sigma_4\bar{\sigma}_6\bar{\sigma}_2}(\qv_{3,4},\qv_{2,8}) \, \Vantisymm_{\bar{\sigma}_4\bar{\sigma}_3\sigma_1\sigma_5}(\qv_{3,4},\qv_{1,7}) \, \Vantisymm_{\sigma_5\bar{\sigma}_2\sigma_6\bar{\sigma}_1}(\qv^+_{2,7},\qv^+_{1,8})\big]}{E_{1,3,4,7}\;E_{2,3,4,8}}
    \\
    + (-1)^{\sigma_1+\sigma_3+\sigma_5+\sigma_6}\;u^2_{7}\,v^2_{12}\; \frac{\Re\big[\Vantisymm_{\bar{\sigma}_2\bar{\sigma}_5\bar{\sigma}_4\bar{\sigma}_1}(\qv_{12,2},\qv_{1,4}) \, \Vantisymm_{\bar{\sigma}_4\bar{\sigma}_3\sigma_1\sigma_6}(\qv_{3,4},\qv_{1,7}) \, \Vantisymm_{\sigma_5\sigma_3\bar{\sigma}_6\bar{\sigma}_2}(\qv_{12,3},\qv_{2,7})\big]}{E_{1,3,4,7}\;E_{2,3,7,12}}
    \\
    - 2\, (-1)^{\sigma_2+\sigma_3+\sigma_4+\sigma_6}\;u^2_{5}\,u^2_{7}\; \frac{\Re\big[\Vantisymm_{\bar{\sigma}_3\bar{\sigma}_1\sigma_2\sigma_5}(\qv_{1,3},\qv_{2,5}) \, \Vantisymm_{\sigma_3\sigma_4\bar{\sigma}_6\bar{\sigma}_1}(\qv_{3,4},\qv_{1,7}) \, \Vantisymm_{\sigma_5\bar{\sigma}_4\sigma_6\bar{\sigma}_2}(\qv^+_{4,5},\qv^+_{2,7})\big]}{E_{1,2,3,5}\;E_{1,3,4,7}}
    \Bigg],
\end{multline}

\begin{multline}
    A_5 = \sum_{\sigma_1\dots\sigma_6} \Bigg[ (-1)^{\sigma_1+\sigma_3+\sigma_4+\sigma_6}\;u^2_{7}\,u^2_{8}\; \frac{\Re\big[\Vantisymm_{\sigma_1\sigma_5\sigma_2\sigma_6}(\qv_{1,7},\qv_{2,8}) \, \Vantisymm_{\sigma_2\sigma_3\bar{\sigma}_6\bar{\sigma}_4}(\qv_{2,3},\qv_{4,8}) \, \Vantisymm_{\bar{\sigma}_3\bar{\sigma}_1\sigma_4\sigma_5}(\qv_{1,3},\qv_{4,7})\big]}{E_{1,3,4,7}\;E_{2,3,4,8}}
    \\
    + \frac{1}{2} (-1)^{\sigma_1+\sigma_4+\sigma_5+\sigma_6}\;u^2_{7}\,v^2_{10}\; \frac{\Re\big[\Vantisymm_{\sigma_1\bar{\sigma}_5\sigma_2\bar{\sigma}_3}(\qv^+_{1,10},\qv^+_{2,3}) \, \Vantisymm_{\bar{\sigma}_3\bar{\sigma}_1\sigma_4\sigma_6}(\qv_{1,3},\qv_{4,7}) \, \Vantisymm_{\sigma_5\sigma_2\bar{\sigma}_6\bar{\sigma}_4}(\qv_{10,2},\qv_{4,7})\big]}{E_{1,3,4,7}\;E_{2,4,7,10}}
    \\
    - (-1)^{\sigma_2+\sigma_3+\sigma_4+\sigma_5}\;u^2_{5}\,u^2_{7}\; \frac{\Re\big[\Vantisymm_{\sigma_2\sigma_3\bar{\sigma}_6\bar{\sigma}_1}(\qv_{2,3},\qv_{1,5}) \, \Vantisymm_{\bar{\sigma}_3\bar{\sigma}_1\sigma_4\sigma_5}(\qv_{1,3},\qv_{4,7}) \, \Vantisymm_{\bar{\sigma}_6\bar{\sigma}_2\bar{\sigma}_5\bar{\sigma}_4}(\qv_{2,5},\qv_{4,7})\big]}{E_{1,2,3,5}\;E_{1,3,4,7}}
    \Bigg],  
\end{multline}

\begin{multline}
    A_6 = \sum_{\sigma_1\dots\sigma_6} \Bigg[ - (-1)^{\sigma_1+\sigma_2+\sigma_4+\sigma_5}\;u^2_{6}\,u^2_{7}\; \frac{\Re\big[\Vantisymm_{\sigma_1\sigma_3\bar{\sigma}_6\bar{\sigma}_4}(\qv_{1,3},\qv_{4,7}) \, \Vantisymm_{\bar{\sigma}_2\bar{\sigma}_1\sigma_4\sigma_5}(\qv_{1,2},\qv_{4,6}) \, \Vantisymm_{\sigma_2\bar{\sigma}_6\sigma_3\bar{\sigma}_5}(\qv^+_{2,7},\qv^+_{3,6})\big]}{E_{1,2,4,6}\;E_{1,3,4,7}}
    \\
    + \frac{1}{8} (-1)^{\sigma_1+\sigma_2+\sigma_3+\sigma_5}\;u^2_{6}\,v^2_{9}\; \frac{\Re\big[\Vantisymm_{\sigma_1\sigma_2\sigma_3\sigma_5}(\qv_{12},\qv_{39}) \, \Vantisymm_{\bar{\sigma}_2\bar{\sigma}_1\sigma_4\sigma_6}(\qv_{1,2},\qv_{4,6}) \, \Vantisymm_{\sigma_4\sigma_6\bar{\sigma}_5\bar{\sigma}_3}(\qv_{4,6},\qv_{3,9})\big]}{E_{1,2,4,6}\;E_{3,4,6,9}}
    \Bigg]  
\end{multline}

\begin{equation}
    A_7 = \sum_{\sigma_1\dots\sigma_6} \frac{1}{8} (-1)^{\sigma_1+\sigma_2+\sigma_4+\sigma_6}\;u^2_{5}\,u^2_{6}\; \frac{\Re\big[\Vantisymm_{\sigma_1\sigma_2\bar{\sigma}_6\bar{\sigma}_4}(\qv_{1,2},\qv_{4,6}) \, \Vantisymm_{\bar{\sigma}_2\bar{\sigma}_1\sigma_3\sigma_5}(\qv_{1,2},\qv_{3,5}) \, \Vantisymm_{\sigma_3\sigma_5\sigma_4\sigma_6}(\qv_{3,5},\qv_{4,6})\big]}{E_{1,2,3,5}\;E_{1,2,4,6}}\,.
\end{equation}
\end{widetext}
%%%%%%%%%%%%%%%%%%%%%%%%%%%%%%%%%%%%%%%%%%%%%%%%%%%%%%%%%%%%%%%%%%%%%%%%%%%%%%%%%%


\begin{thebibliography}{99}
\bibitem{Chamel2008} N.~Chamel and P.~Haensel,
  %``Physics of Neutron Star Crusts,''
  Living Rev.\ Rel.\ \textbf{11} (2008), 10 [doi:10.12942/lrr-2008-10].
\bibitem{Anderson1975} P.~W.~Anderson and N.~Itoh, Nature (London) \textbf{256}, 25 (1975).
\bibitem{Chamel2012} N.~Chamel, Phys. \ Rev. \ C \textbf{85}, 035801 (2012).
\bibitem{Andersson2012} N.~Andersson, K.~Glampedakis, W.~C.~G.~Ho, and C.~M.~Espinoza, Phys.\ Rev.\ Lett.\ \textbf{109}, 241103 (2012).
\bibitem{Martin2016} N.~Martin and M.~Urban, Phys.\ Rev.\ C \textbf{94}, 065801 (2016).
\bibitem{Wlazlowski2016} G.~Wlaz\l{}owski, K.~Sekizawa, P.~Magierski, A.~Bulgac, and M.~M.~Forbes,
Phys.\ Rev.\ Lett.\ \textbf{117}, 232701 (2016).
\bibitem{Yakovlev2004} D.~G.~Yakovlev and C.~J.~Pethick, Ann.\ Rev.\ Astron.\
  Astrophys.\ \textbf{42}, 169 (2004).
\bibitem{Migdal1959} A.~B.~Migdal, Zh.\ Eksp.\ Teor.\ Fiz.\  \textbf{37}, 249 (1959).
\bibitem{Martin2015} N.~Martin and M.~Urban, Phys.\ Rev.\ C \textbf{92}, 015803 (2015).
\bibitem{Ramanan2020} S.~Ramanan and M.~Urban, 
  %``Pairing in pure neutron matter,''
  Eur.\ Phys.\ J.\ ST \textbf{230}, 567 (2021).
\bibitem{Dong2013} J.~M.~Dong, U.~Lombardo and W.~Zuo,
  %``$^3PF_2$ pairing in high-density neutron matter,''
  Phys.\ Rev.\ C \textbf{87}, 062801(R) (2013).
\bibitem{Maurizio2014} S.~Maurizio, J.~W.~Holt, and P.~Finelli,
Phys.\ Rev.\ C \textbf{90}, 044003 (2014). 
\bibitem{Srinivas2016} S.~Srinivas and S.~Ramanan, Phys.\ Rev.\ C \textbf{94}, 064303 (2016).
\bibitem{Drischler2017} C.~Drischler, T.~Kr\"uger, K.~Hebeler and A.~Schwenk,
  %``Pairing in neutron matter: New uncertainty estimates and three-body forces,''
  Phys.\ Rev.\ C \textbf{95}, 024302 (2017).
\bibitem{Papakonstantinou2017} P.~Papakonstantinou and J.~W.~Clark,
  %``Three-Nucleon Forces and Triplet Pairing in Neutron Matter,''
  J.\ Low Temp.\ Phys.\ \textbf{189}, 361 (2017).
\bibitem{Baker1999} G.~A.~Baker, Phys.\ Rev.\ C \textbf{60}, 054311 (1999).
\bibitem{CalvaneseStrinati2018} G.~Calvanese Strinati, P.~Pieri, G.~R\"opke, P.~Schuck, and
  M.~Urban, Phys.\ Rep.\ \textbf{738}, 1 (2018).  
\bibitem{Urban2021} M.~Urban and S.~Ramanan, 
  %``Low-momentum interactions for ultracold Fermi gases,'' 
    Phys. Rev. A \textbf{103}, 063306 (2021).
\bibitem{Tichai2018} A.~Tichai, P.~Arthuis, T.~Duguet, H.~Hergert, V.~Som\`a, and
  R.~Roth,
  %``Bogoliubov many-body perturbation theory for open-shell nuclei,''
  Phys.\ Lett.\ B \textbf{786}, 195 (2018).
\bibitem{Bogner2002} S.~Bogner, T.~T.~S.~Kuo, L.~Coraggio, A.~Covello, and N.~Itaco,
Phys.\ Rev.\ C \textbf{65}, 051301(R) (2002).
\bibitem{Bogner2003} S.~Bogner, T.~T.~S.~Kuo and A.~Schwenk, Phys. Rep. \textbf{386}, 1 (2003).
\bibitem{Bogner2007} S.~K.~Bogner, R.~J.~Furnstahl, S.~Ramanan, and A.~Schwenk, 
  Nucl.\ Phys.\ A  \textbf{784}, 79 (2007).
%\bibitem{Sakurai} J. J. Sakurai, \textit{Modern Quantum Mechanics}, Revised Edition (Addison-Wesley, Reading, 1994).
\bibitem{Bogner2010} S.~K.~Bogner, R.~J.~Furnstahl and A.~Schwenk,
  %``From low-momentum interactions to nuclear structure,''
  Prog.\ Part.\ Nucl.\ Phys.\ \textbf{65} 94, (2010).
\bibitem{Schwenk2003} A.~Schwenk, B.~Friman and G.~E.~Brown, Nucl.\ Phys.\ A \textbf{713}, 191 (2003).
\bibitem{Ramanan2018} S.~Ramanan and M.~Urban, 
  %``Screening and antiscreening of the pairing interaction in low-density 
  %neutron matter,'' 
  Phys.\ Rev.\ C \textbf{98}, 024314 (2018).
\bibitem{Coraggio2013} L.~Coraggio, J.~W.~Holt, N.~Itaco, R.~Machleidt and
  F.~Sammarruca,
 %``Reduced regulator dependence of neutron-matter predictions with perturbative 
 %chiral interactions,''
  Phys.\ Rev.\ C \textbf{87}, 014322 (2013).
\bibitem{FetterWalecka} A.~L.~Fetter and J.~D.~Walecka, \textit{Quantum Theory of
  Many-Particle Systems} (McGraw-Hill, New York, 1971).  
\bibitem{Varshalovich} D.~A.~Varshalovich, A.~N.~Moskalev, and V.~K.~Khersonskii,
  \textit{Quantum Theory of Angular Momentum} (World Scientific, Singapore, 1988).
\bibitem{Wiringa1995}  R.~B.~Wiringa, V.~G.~J.~Stoks, and R.~Schiavilla, 
  Phys.\ Rev.\ C \textbf{51}, 38 (1995).
\bibitem{Entem2003} D.~R.~Entem and R.~Machleidt, Phys.\ Rev.\ C \textbf{68}, 041001(R)
  (2003).
\bibitem{Wellenhofer2021} C.~Wellenhofer, C.~Drischler, and A.~Schwenk,
  %``Effective field theory for dilute Fermi systems at fourth order,''
  Phys.\ Rev.\ C \textbf{104}, 014003 (2021).
\bibitem{Ramanan2007} S.~Ramanan, S.~K.~Bogner and R.~J.~Furnstahl, Nucl.\ Phys.\ A \textbf{797}, 81 (2007).  
\bibitem{Gezerlis2010} A.~Gezerlis and J.~Carlson,
  %``Low-density neutron matter,''
  Phys.\ Rev.\ C \textbf{81}, 025803 (2010).
\bibitem{Gandolfi2022}
  S.~Gandolfi, G.~Palkanoglou, J.~Carlson, A.~Gezerlis and K.~E.~Schmidt,
  %``The 1S0 Pairing Gap in Neutron Matter,''
  Condens.\ Mat.\ \textbf{7}, 19 (2022).
\bibitem{Lovato2022}
  A.~Lovato, I.~Bombaci, D.~Logoteta, M.~Piarulli and R.~B.~Wiringa,
  %``Benchmark calculations of infinite neutron matter with realistic two- and three-nucleon potentials,''
  Phys.\ Rev.\ C \textbf{105}, 055808 (2022).
\bibitem{Hebeler2010}
  K.~Hebeler and A.~Schwenk,
  %``Chiral three-nucleon forces and neutron matter,''
  Phys.\ Rev.\ C \textbf{82}, 014314 (2010).
\bibitem{Nogga2004} A.~Nogga, S.~K.~Bogner, and A.~Schwenk, 
  % ``Low-momentum interaction in few-nucleon systems,''
  Phys.\ Rev.\ C \textbf{70}, 061002(R) (2004).
  \bibitem{Hergert2016}
H.~Hergert, S.~K.~Bogner, T.~D.~Morris, A.~Schwenk and K.~Tsukiyama,
%``The In-Medium Similarity Renormalization Group: A Novel Ab Initio Method for Nuclei,''
Phys. Rep. \textbf{621} 165, (2016).
\bibitem{Schrieffer} J.~R.~Schrieffer, \textit{Theory of Superconductivity} (Benjamin, New York, 1964).
\end{thebibliography}
\end{document}